\documentclass[pra,showpacs,preprintnumbers,amsmath,amssymb]{revtex4}

\usepackage{bm}
\usepackage{amsmath}
\usepackage[dvips]{graphicx}
\newcommand{\del}[1]{}

\begin{document}
\bibliographystyle{apsrev}

\title{Virtual light-by-light scattering and the $g$ factor of a bound electron}

\author{R.N.Lee}
\email{R.N.Lee@inp.nsk.su}
\author{A.I. Milstein}
\email{A.I.Milstein@inp.nsk.su}
\author{I.S. Terekhov}
\email{I.S.Terekhov@inp.nsk.su}
\affiliation{Budker Institute of Nuclear Physics, 630090
Novosibirsk, Russia}

\author{S.G. Karshenboim}
\email{sek@mpq.mpg.de}
\affiliation{D. I. Mendeleev Institute for Metrology (VNIIM),
    St. Petersburg 198005, Russia\\
Max-Planck-Institut f\"ur Quantenoptik, 85748 Garching, Germany}

\date{\today}

\begin{abstract}
The contribution of the light-by-light diagram to the $g$ factor
of electron and muon bound in Coulomb field is obtained. For
electron in a ground state, our results are in good agreement with
the results of other authors obtained  numerically for large $Z$.
For relatively small $Z$ our results have essentially higher
accuracy as compared to the previous ones. For muonic atoms, the
contribution is obtained for the first time with the high accuracy
in whole region of $Z$.
\end{abstract}

\maketitle

\section{Introduction}

\begin{figure}[h]
 \includegraphics[height=2.5 cm]{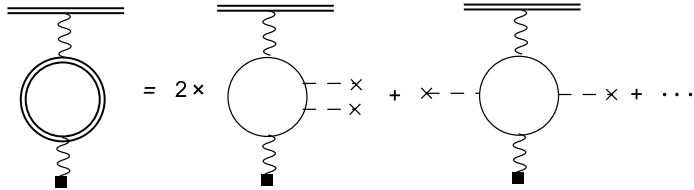}
\caption{The diagram corresponding to the magnetic-loop contribution
to the $g$ factor of a bound electron and first non-vanishing terms
of expansion of this loop with respect to the Coulomb field. Double
line denotes the electron propagator and the wave function in a
Coulomb field, the dashed line with the cross denotes the Coulomb
field, the wavy line with the square denotes the external
homogeneous magnetic field, the internal wavy line corresponds to
the photon propagator.}
 \label{fig:magneticloop}
\end{figure}

The progress in experimental investigations of the $g$ factor of a
bound electron \cite{eions} and muon \cite{muhe,muonic} in ions
stimulated intensive theoretical investigation of various
contributions to this quantity. The contributions of self-energy,
vacuum polarization, and nuclear effects have been considered
\cite{selfen,vp,recoil,Beier2000,JETPVP,PLA,MilKarsh2002}. An
essential part of the theoretical uncertainty has been related to
the contribution of the vacuum polarization of an external
homogeneous magnetic field in the electric field of atom (so-called
the ``magnetic-loop'' contribution). The corresponding diagram is
shown in Fig. \ref{fig:magneticloop}. In this diagram, double line
in the fermion loop corresponds to the electron propagator in the
Coulomb field. Note that the contribution of the free electron loop
to the vacuum polarization of a homogeneous magnetic field vanishes
due to the gauge invariance. The first non-vanishing terms of
expansion with respect to the Coulomb field shown in Fig.
\ref{fig:magneticloop} is the contribution of virtual light-by-light
scattering with one of the quanta corresponding to the external
magnetic field. The results of numerical calculations of the
magnetic-loop contribution, which take into account all orders of
the parameter $Z\alpha$ ($Z$ is the nuclear charge number,
$\alpha=e^2$ is the fine-structure constant, $\hbar=c=1$), are
presented in Ref. \cite{Beier2000}. At present, the most accurate
experimental data are obtained in the region of medium $Z$.
Unfortunately, in this region the uncertainty of the results of Ref.
\cite{Beier2000} is very big, being, e.g., $100\%$ for $Z=12$. In
Ref. \cite{MilKarsh2002}, the leading in $Z\alpha$ magnetic-loop
contribution to the $g$ factor of an electron in $S$ state of a
hydrogen-like ion has been derived. It reads
\begin{equation}\label{eq:aZa5}
\frac{\Delta g_0}{g_0}=\frac{\Delta g_0}2=\frac{7\alpha(Z\alpha)^5}{432n^3}\,,
\end{equation}
where $g_0$ is the Land\'e factor equal to two for $S$ state. One
can compare this correction with the result of \cite{Beier2000} for
rather large $Z$ where the accuracy of the numerical calculation is
reasonable. This comparison shows the noticeable difference which
can be attributed to the contribution of the next-to-leading terms
in $Z\alpha$-expansion, starting from $\alpha (Z\alpha)^6$. Since
the numerical factor in Eq. (\ref{eq:aZa5}) is very small ($\sim
1/30$), the next-to-leading terms could give a noticeable
contribution to the $g$ factor even at small $Z$, if the
corresponding numerical factor is of order of unity.

In the present paper, we generalize Eq. (\ref{eq:aZa5}) to the case
of arbitrary bound electron state. We also calculate the
next-to-leading contribution of magnetic loop to the $g$ factor of
the electron in arbitrary state (or the magnetic moment of the
electron in this state). It has the form $\Delta
g_1=\alpha(Z\alpha)^6(a_1\ln(1/Z\alpha)+a_2)$, where $a_{1,2}$ are
some constants and $a_1$ is not zero only for $S$ states. In order
to calculate this contribution, it is sufficient to take into
account the diagrams of virtual light-by-light scattering and use
the nonrelativistic wave functions of the bound electron. Comparison
of the correction $\Delta g_0+\Delta g_1$ for $1S_{1/2}$ state with
the results of \cite{Beier2000} shows that the account of $\Delta
g_1$ does not provide good agreement for relatively small $Z\sim
30$, where the numerical calculations were performed with sufficient
accuracy. Thus, for such $Z$ it is necessary to take into account
next terms in $Z\alpha$. These terms have two different origins.
First, they come from the relativistic corrections to the wave
function of a bound electron. Next, they come from the higher-order
contributions to the electron loop. Note that the diagram in Fig.
\ref{fig:magneticloop} can be interpreted as the contribution of the
scattering of the magnetic quantum in a Coulomb field (virtual
Delbr\"uck scattering) to the $g$ factor. It is known that the
Coulomb corrections to the Delbr\"uck amplitude for momentum of
quantum $q\lesssim m$ ($m$ is the electron mass) are numerically
small even for large $Z$ \cite{delb1,delb2}. In contrast, the
account of the corrections to the wave function is very important,
starting from relatively small $Z$. We calculate the correction
$\Delta g$ using the relativistic wave function and the leading
approximation for the electron loop. As a result we have obtained
good agreement with the numerical data of \cite{Beier2000} even for
very large $Z$ (difference is $4\%$ for $Z=92$). Using such
approach, we have calculated the corresponding correction of the
electron loop to the $g$ factor of a bound muon.

\section{General relations}
Let us consider the amplitude $T$ of interaction of homogeneous
magnetic field $\bm B$ with the electron bound in a hydrogenlike
ion. In the zero approximation, it reads (see, e.g.,
\cite{rose1961})
\begin{equation}
\label{eq:TM0} T^{(0)}=e\int \frac{d\bm k}{(2\pi)^3} {\bm A}_{\bm k}
\cdot {\bm j^*_{\bm k}}=\frac{e\kappa \bm B\cdot \langle\bm J
\rangle}{J(J+1)} \int_0^\infty dr r^3 f_1(r)f_2(r)\,,
\end{equation}
where $\bm j_{\bm k}$ is the Fourier transform of the electron
current $\bm j(\bm r) =\bar\psi(\bm{r}) \bm{\gamma} \psi
(\bm{r})$, the wave function $\psi$ has the form
\begin{equation}
\label{Dirac} \psi({\bm r})= \begin{pmatrix}
f_1(r)\Omega\\
if_2(r)\tilde{\Omega}
\end{pmatrix}
\, ,
\end{equation}
$\Omega$ is the spherical spinor \cite{BLP} with the angular
momentum $J$ and orbital momentum $L$, $\tilde{\Omega}=-({\bm
\sigma}\cdot{\bm n})\Omega$, $\kappa=(J+1/2)\mbox{sign}(L-J)$. In
Eq. (\ref{eq:TM0}) we have used the relation
\begin{equation}
\label{BviaA} i \bm k\times\bm A_{\bm k}=(2\pi)^3 \delta(\bm k)
\bm B\,.
\end{equation}
Note that a sign of $T^{(0)}$ is opposite to that of Hamiltonian.
Substituting the radial wave functions $f_1(r)$ and $f_2(r)$ for the
Coulomb field (see, e.g., \cite{BLP}), we obtain for the arbitrary
bound state
\begin{eqnarray}
\label{eq:TM0r} &\displaystyle T^{(0)}=\frac{e\bm B\cdot
\langle\bm
J\rangle}{2m}g\,,\nonumber\\
&\displaystyle
g=\frac{2\kappa}{1-4\kappa^2}\left(1-\frac{2\kappa\varepsilon}m\right)=
\frac{2\kappa }{1-4\kappa^2}
\left(1-\frac{2\kappa}{\sqrt{1+(Z\alpha)^2/(\gamma+n_r)^2}}\right)\,,
\end{eqnarray}
where $n_r$ is the radial quantum number, $\varepsilon$ is the
binding energy, and $\gamma=\sqrt{\kappa^2-(Z\alpha)^2}$. The
particular cases of this formula obtained earlier are presented in
\cite{rose1961}. In the non-relativistic approximation
($Z\alpha\ll 1$), Eq. (\ref{eq:TM0}) turns to
\begin{equation}
\label{eq:nrTM0} T_0^{(0)}=\frac{e\bm B\cdot \langle\bm
J\rangle}{2m}g_0\,,\quad g_0=\frac{2\kappa}{2\kappa+1}\,.
\end{equation}

We now pass to the calculation of the amplitude $T^{(1)}$
corresponding to the diagram shown in Fig. \ref{fig:magneticloop}.
It has the form
\begin{equation}\label{TM1}
{\cal T}^{(1)}=e\int \frac{d\bm k}{(2\pi)^3}\int \frac{d\bm
q}{(2\pi)^3}\frac{4\pi}{\bm q^2}\, A_{\bm k}^i\,{\cal
M}^{i\,l}\,j_{\bm q}^{l*}\,,
\end{equation}
where the amplitude ${\cal M}^{i\,l}$ of the virtual Delbr\"uck
scattering in the case $k\ll m$ has the form following from the
gauge invariance
\begin{eqnarray}
\label{eq:T_il} {\cal
M}^{i\,l}=\frac{\alpha}{m^3}[\delta^{i\,l}(\mathbf{k}\cdot
\mathbf{q})-q^i k^l]\,{\cal F}(q/m,Z\alpha)\,.
\end{eqnarray}
Note that $\cal F$ is even function of $Z\alpha$. In the leading in
$Z\alpha$ approximation (contribution of light-by-light scattering),
\begin{equation}
\label{Za2F}
{\cal F}(q/m, Z\alpha)=(Z\alpha)^2 F(q/m)\,,
\end{equation}
with $F(0)=7/1152$, see Ref. \cite{MilKarsh2002}. From Eqs.
(\ref{BviaA}), (\ref{TM1}), (\ref{eq:T_il}), and (\ref{Za2F}) we
obtain
\begin{equation}\label{eq:TM1f}
T^{(1)}=e \frac{4\kappa \alpha (Z\alpha)^2 \bm B\cdot \langle\bm
J\rangle}{\pi m^3 J(J+1)}\int\limits_0^\infty dq
F(q/m)\int\limits_0^\infty dr r f_1(r)f_2(r) \left(\frac{\sin
qr}{qr}-\cos qr\right)
\end{equation}

Using the relation ${\cal M}^{ii}=[2\alpha (Z\alpha)^2(\bm k\cdot\bm
q)/m^3]F(q/m)$, following from Eq. (\ref{eq:T_il}), and the gauge
invariance of the light-by-light scattering amplitude, we can
represent $F(q/m)$ in the following form
\begin{eqnarray}\label{F1}
F(q/m) &=& \frac {m^3}{2\pi} \int \frac{d\bm Q}{\bm Q^2 (\bm q-
\bm Q)^2}
\frac{\bm q\cdot (\nabla_{\bm k} M)|_{\bm k=0}}{\bm q^2}\,,\nonumber\\
M&=& 2i\int \frac{d^4 p}{(2\pi)^4}
\mbox{Sp}\biggl\{G(p)\gamma^iG(p-k)\gamma^0 \biggl[G(p+Q-q)\gamma^i
G(p+Q)\gamma^0\nonumber\\
&&+G(p+Q-q)\gamma^0G(p-q)\gamma^i+G(p-Q-k)\gamma^0G(p-q)\gamma^i\biggr]\biggr\}\,,
\end{eqnarray}
where $G(p)=[\hat p-m]^{-1}$ is a free electron propagator.
Straightforward calculation leads to the representation of the
function $F$ in the form of two-fold integral with respect to the
Feynman parameters. Resulting formulas being rather cumbersome are
not presented here explicitly. For $x=q/m\ll 1$, the first two
terms of expansion of the function $F(x)$ have the form
\begin{eqnarray}\label{eq:Fas}
F(x) = \frac{7}{1152}(1+\frac{8}{35}x)\,.
\end{eqnarray}
The first term in this formula agrees with the result of Ref.
\cite{MilKarsh2002}. For $x\gg 1$, the asymptotics of the function
$F(x)$ reads
\begin{eqnarray}\label{eq:Fas2}
F(x) = \frac1{2x^3}\,.
\end{eqnarray}

\begin{figure}[h]
\vspace{40pt} \centering \setlength{\unitlength}{1pt}
\includegraphics[height=140\unitlength,keepaspectratio=true]{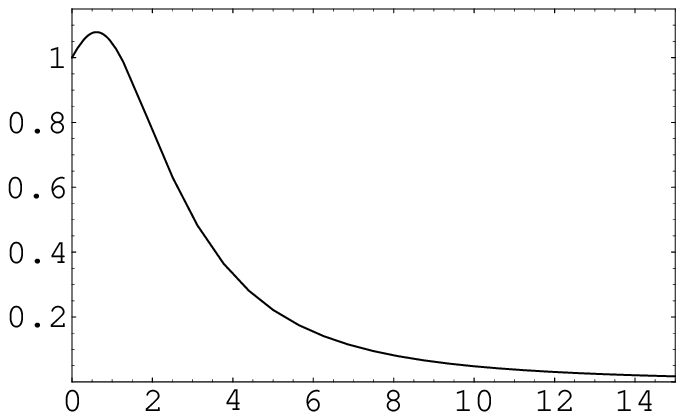}
\begin{picture}(0,0)(0,0)
 \put(-110,-4){\large $x$}
 \put(-240,50){\rotatebox{90}{$F(x)/F(0)$}}
 \end{picture}
\caption{The ratio $F(x)/F(0)$ as a function of $x=q/m$ .}
\label{Fig_f}
\end{figure}

For arbitrary $x$, we performed the numerical tabulation of the
function $F(x)$. The result is shown in Fig.~\ref{Fig_f} and in Table
\ref{tableF}.

\section{Correction to \lowercase{{\large$g$}} factor at small $Z\alpha$}

In order to obtain the leading term of expansion in $Z\alpha$ of the
amplitude ${\cal T}^{(1)}$, it is sufficient to use  Eq.
(\ref{eq:TM1f}) with the substitution $F(q/m)\to F(0)$ and the wave
functions taken in the nonrelativistic approximation. In this
approximation $f_1(r)$ coincides with $R(r)$, the radial part of the
nonrelativistic wave function, and
\begin{equation}\label{eq:fg_nr}
f_2(r)=\frac1{2m}\left(R'(r)+\frac{1+\kappa}{r}R(r)\right)\,.
\end{equation}

The correction $\Delta g$ to the Land\'e factor is determined by
the relation
\begin{equation}
\frac{\Delta g}{g_0} =\frac{{\cal T}^{(1)}}{T^{(0)}_0}\,.
\end{equation}

Taking in Eq.(\ref{eq:TM1f}) the integral over $q$, and then over
$r$, we obtain  the leading contribution $\Delta g_0$ for the
arbitrary state
\begin{equation}\label{eq:T10}
\frac{\Delta g_0}{g_0}= \frac{7\alpha(Z\alpha)^5}{144
n^3(2L+1)\kappa(2\kappa-1)}=\frac{7\alpha(Z\alpha)^5}{288
n^3J(J+1)(2J+1)}\,,
\end{equation}
where $n=n_r+|\kappa|$ is a principal quantum number. For $S$
states ($L=0$, $\kappa=-1$), this result is in agreement with Eq.
(\ref{eq:aZa5}) obtained in Ref. \cite{MilKarsh2002}.

The relativistic corrections to the wave function as well as the
corrections to the magnetic loop have the relative magnitude
$(Z\alpha)^2$. Therefore, the term $\Delta g_1$ of the order
$\alpha (Z\alpha)^6$  can also be obtained with the use of the
nonrelativistic wave functions and magnetic loop in the leading
approximation (light-by-light scattering diagrams). For $L\neq 0$,
it is sufficient to substitute the second term of expansion of $F(x)$,
 see Eq. (\ref{eq:Fas}), in Eq. (\ref{eq:TM1f}). Then we obtain
\begin{equation}
\frac{\Delta g_1}{g_0}= \frac{2\alpha(Z\alpha)^6}{45\pi n^3
(2L+1)(2\kappa-1)^2}\left(\frac{3}{L(L+1)}-\frac1{n^2}\right)\,.
\end{equation}
For $S$ states, calculation of $\Delta g_1$ is more complicated. For
$nS$ state, $f_1(r)f_2(r)=(\pi/m)\rho'_n(r)$, where $\rho_n(r)$ is
the electron density in the nonrelativistic approximation.
Substitution of Eq.(\ref{eq:Fas}) in Eq.(\ref{eq:TM1f}) leads to
logarithmic divergence. Therefore, it is convenient to split the
region of integration over $r$ in Eq. (\ref{eq:TM1f}) into two:
$[0,r_0]$ and $[r_0,\infty)$ with $1/m\ll r_0\ll 1/(mZ\alpha)$. In
the first region, we can replace $\rho'(r)$ by $\rho'(0)$ and take
the integral over $r$. In the second region, we can use the
expansion (\ref{eq:Fas}) and take the integral over $q$. Sum of
these two contributions, as it should be, is independent of $r_0$.
The final result reads
\begin{eqnarray}\label{eq:T11s}
\frac{\Delta g_1}{g_0}&=& \frac{4\alpha(Z\alpha)^6}{135\pi
n^3}\left(\ln\frac1{Z\alpha}-a-b_n\right)\,,\nonumber\\
a&=&-\frac12+\frac{35}{8}\int\limits_0^\infty dx \ln x
F''(x)\approx2.6\,,\nonumber\\
b_n&=&-C+\frac1{\rho'_n(0)}\int\limits_0^\infty dr \ln(mZ\alpha r)
 \rho''_n(r)\,,
\end{eqnarray}
where $C=0.577...$ is the Euler constant. For each $n$, the
coefficient $b_n$ can be easily calculated so that $b_1=\ln
2\approx 0.693$, $b_2=5/8=0.625$, $b_3=55/54+\ln2/3\approx0.613$,
$b_\infty= C+\ln 2-2/3\approx0.604$.

\section{Correction to \lowercase{{\large$g$}} factor at $Z\alpha\sim 1$}

\begin{figure}[h]
 \includegraphics[height=4.5 cm]{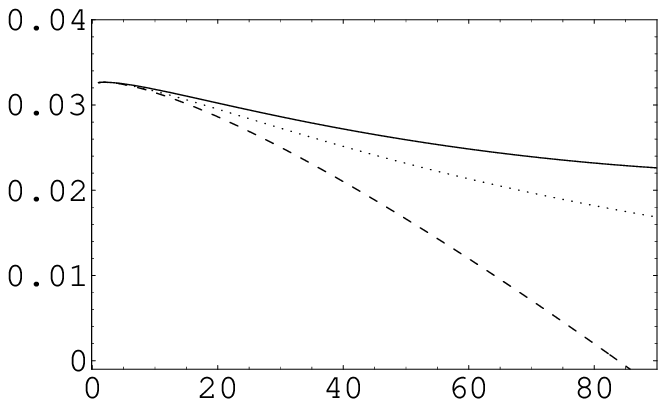}
 \begin{picture}(0,0)(0,0)
 \put(-95,-4){\large $Z$}
 \put(-220,50){\rotatebox{90}{$\Delta g/[\alpha (Z\alpha)^{5}]$}}
 \end{picture}
\caption{The correction $\Delta g$ for  $1S_{1/2}$. Solid curve:
the exact result; dashed curve: $\Delta g_0+\Delta g_1$; dotted
curve:  $\Delta g_{nr}$.}
 \label{Fig2}
\end{figure}

As it was pointed out in the Introduction, the sum $\Delta
g_0+\Delta g_1$ gives a good approximation to $\Delta g$ only for
small $Z$. For intermediate $Z$, it is necessary to account for
the next terms in $Z\alpha$. The largest corrections are due to
the significant difference between the relativistic wave function
and the nonrelativistic one already at intermediate $Z$. At the
same time, the difference between the function $\cal F$ and its
leading approximation $(Z\alpha)^2F$ results in the corrections
which are numerically small even for large $Z$. Using the
numerical results for $F(x)$ and the relativistic wave functions,
we have performed the tabulation of $\Delta g$ for various $Z$,
using $T^{(1)}$ from Eq. (\ref{eq:TM1f}) as an approximation to
${\cal T}^{(1)}$. The results of this tabulation for $1S_{1/2}$,
$2S_{1/2}$, and $2P_{1/2}$ states are presented in Table
\ref{table1}. For $1S_{1/2}$ state, we also present the
contribution of the first two terms of expansion in $Z\alpha$,
Eqs. (\ref{eq:T11s}), (\ref{eq:T10}), and the correction $\Delta
g_{nr}$ obtained with the use of nonrelativistic wave functions.
The results for $1S_{1/2}$ are also shown in Fig.\ref{Fig2}.

For $Z<10$, both $\Delta g_0+\Delta g_1$ and $\Delta g_{nr}$
coincide with $\Delta g$ with accuracy better than one percent.
The difference grows with $Z$, reaching $10\%$ at $Z\sim 30$ for
$\Delta g_0+\Delta g_1$ and at $Z\sim 50$ for $\Delta g_{nr}$.

In Table \ref{table1}, we also show the results of numerical
tabulation from Ref. \cite{Beier2000} for $1S_{1/2}$ state. For
$30<Z<70$, our result for $\Delta g$ agrees with that obtained in
Ref. \cite{Beier2000} within $1\div2\%$ percent. The difference
between these two results for $Z<30$ is due to poor accuracy of
the numerical results of Ref. \cite{Beier2000}. For $Z>70$ the
difference increases and becomes $8\%$ for $Z=92$. This difference
corresponds to the contribution of next-to-leading terms in
magnetic loop, that was taken into account in Ref.
\cite{Beier2000} and omitted in our paper. Thus, the effect of
these terms is small in a wide region of $Z$, while the
relativistic effects in the wave function become important already
at relatively small $Z$.

\section{The correction \lowercase{{\large$\Delta g$}} for muonic atoms}

The correction $\Delta g$ to the $g$ factor of a bound muon due to
the electron magnetic loop can be obtained from Eq. (\ref{eq:TM1f})
with $f_1(r)$ and $f_2(r)$ being the wave functions of the muon. The
asymptotics of $\Delta g$ for $\mu Z\alpha/(mn^2)\approx 1.5
Z/n^2\gg 1$ ($\mu$ is the muon mass) can be calculated as follows.
We split the region of integration over $q$ in Eq. (\ref{eq:TM1f})
into two: $[0,q_0]$ and $[q_0,\infty)$ with $m\ll q_0\ll \mu
Z\alpha/n^2$. In the first region we can replace $[(qr)^{-1}\sin
qr-\cos qr]$ by $(qr)^2/3$ and take the integral over $r$. In the
second region we can use the asymptotics Eq. (\ref{eq:Fas2}) and
take the integral over $q$. Summing these two contributions, we
obtain
\begin{eqnarray}\label{eq:mu_as}
\Delta g_{as}&=&g\frac{2\alpha(Z\alpha)^2}{3\pi}\left[\ln
(\mu Z\alpha/m)-A-B\right]\,,\nonumber\\
A&=&2\int_0^\infty dy \ln y \partial_y\,(y^3F(y))\approx 2.24\,,\nonumber\\
B&=&C-\frac43-\frac{4}{Z\alpha(1-2\kappa\varepsilon/\mu)}\int_0^\infty
dx x^3\, {\tilde f}_1(x){\tilde f}_2(x) \ln x\,,\nonumber\\
{\tilde f}_1(x)&=&(\mu Z\alpha)^{-3/2} f_1(x/\mu Z\alpha)\,,\quad
{\tilde f}_2(x)=(\mu Z\alpha)^{-3/2} f_2(x/\mu Z\alpha)\,,
\end{eqnarray}
$g$ is defined in Eq. (\ref{eq:TM0r}). For $1S_{1/2}$ state we
obtain
\begin{equation}
g=\frac23(1+2\gamma)\,, \quad B=C-\frac43+\psi(2\gamma+2)-\ln
2\,,\quad \gamma=\sqrt{1-(Z\alpha)^2}\,.
\end{equation}
For $n=n_r+|\kappa|\gg 1$, we have
\begin{equation}
g=g_0=\frac{2\kappa}{2\kappa+1}\,,\quad B=C+\ln (n^2/2)\,.
\end{equation}

The formula (\ref{eq:mu_as}) can be interpreted as follows. In Ref.
\cite{MilYelkh1989} the logarithmic contribution of the electron
vacuum polarization to the magnetic moment of a heavy nucleus was
calculated. The result obtained has the form
\begin{eqnarray}\label{eq:gnucl}
\frac{\Delta g}g&=&\frac{2\alpha(Z\alpha)^2 H(Z\alpha)}{3\pi}\ln
(1/mR_{nucl})\,,
\end{eqnarray}
where $R_{nucl}$ is the nuclear radius, $R_{nucl}\ll 1/m$. The
coefficient $(Z\alpha)^2H(Z\alpha)$ was calculated exactly in
$Z\alpha$, i.e., with the account of all Coulomb corrections to the
electron loop. The function $H(Z\alpha)$ tends to unity when
$Z\alpha\to 0$, and significantly differs from unity only for very
large $Z$. Large logarithm $\ln (1/mR_{nucl})$ in Eq.
(\ref{eq:gnucl}) appears as a result of integration over distance
$r$ in the region $R_{nucl}\ll r\ll 1/m$. We can consider the muonic
atom as some nucleus with the effective radius $R_{nucl}\sim n^2/\mu
Z\alpha$. In the case $\mu Z\alpha/(mn^2)\gg 1$, we have
$R_{nucl}\ll 1/m$. Substituting this radius into Eq.
(\ref{eq:gnucl}) and replacing $H(Z\alpha)\to 1$ (that corresponds
to the contribution of light-by-light scattering), we obtain the
logarithmically amplified term in Eq. (\ref{eq:mu_as}). Note that
the coefficient $n^2$ in $R_{nucl}$ corresponds to the asymptotics
of $B$ in Eq. (\ref{eq:mu_as}) at $n\gg 1$. Strictly speaking, the
charge of such effective nucleus is $Z-1$, but not $Z$. However,
under the condition $\mu Z\alpha/(mn^2)\approx 1.5 Z/n^2\gg 1$, this
difference is not important.

In Table \ref{tablemu}, we present $\Delta g$ for $1S_{1/2}$ state
of muonic atom calculated for arbitrary $Z$. For comparison, we
present also asymptotics Eq. (\ref{eq:mu_as}). As it should be, the
accuracy of asymptotics (\ref{eq:mu_as}) increases with $Z$ being
$4\%$ for $Z=40$ and $1\%$ for $Z=92$.

In summary, we found higher-order magnetic loop corrections to the
bound $g$ factor in order $\alpha (Z\alpha)^6$ for arbitrary
state. Despite a small coefficient in the leading term of order
$\alpha (Z\alpha)^5$, and the logarithmic enhancement of the
higher-order contribution, the leading term still dominate for
$Z=6$ and $Z=8$, important for experiment. Previously used
numerical results show a certain underestimation of the
magnetic-loop contribution for $Z<20$. Theoretical description of
this contribution presented  in this paper is more reliable. The
difference of less than a few percent between our analytic results
and the numerical calculations of Ref. \cite{Beier2000} at high
$Z$ ($80\div 90$) shows that the contribution of the higher-order
terms in magnetic loop may be safely neglected for $Z\lesssim 50$.
We also calculated the correction $\Delta g$ for the bound muon,
and its behavior is very peculiar. All known contributions to the
bound $g$ factor scale as $n^{-2}$ or $n^{-3}$. The correction
found in this paper does not contain such a strong suppression
factor. This correction is a dominant bound state QED correction
for a bound muon, which even for $1S_{1/2}$ state supersedes the
free vacuum polarization term \cite{JETPVP}. The results obtained
significantly diminish the uncertainty of the theoretical
predictions for the $g$ factor of a bound particle.

This work was supported in part by Russian Science Support
Foundation, RFBR Grants Nos. 03-02-16510 and 03-02-16843.

\newpage

\begin{table}[h]
\begin{tabular}{|cc|cc|cc|cc|cc|cc|}
\hline
$x$&$F(x)/F(0)$&$x$&$F(x)/F(0)$&$x$&$F(x)/F(0)$&$x$&$F(x)/F(0)$&$x$&$F(x)/F(0)$&$x$&$F(x)/F(0)$\cr\hline
$0.$&$1. $&                $ 0.42 $&$ 1.07 $ &   $ 1.5 $&$0.927$& $
5. $&$ 0.222 $ &                        $ 26 $&$ 3.81\,
\times\,{10}^{-3} $ &     $ 128 $&$ 3.75\, \times \, {10}^{-5} $\cr
$ 0.05$&$ 1.01$&           $ 0.44 $&$ 1.07 $&    $ 1.6 $&$ 0.897 $ &
$ 5.5 $&$ 0.184$&                        $ 28 $&$ 3.1\, \times \,
{10}^{-3} $ &    $ 144 $&$ 2.66\, \times \, {10}^{-5} $ \cr $ 0.1
$&$ 1.02 $ &        $0.46 $&$ 1.07 $ &     $ 1.7 $&$ 0.867 $ &    $
6. $&$ 0.154 $ &                       $ 30$&$2.55\, \times \,
{10}^{-3} $ &      $ 160 $&$1.94\,\times \, {10}^{-5} $ \cr $ 0.15
$&$ 1.03 $ &        $ 0.48 $&$ 1.08 $ &   $ 1.8$&$0.837 $ &      $
6.5 $&$ 0.13 $ &                       $ 32 $&$ 2.12\,
\times\,{10}^{-3} $ &      $ 176 $&$ 1.46\, \times\,{10}^{-5} $ \cr
$ 0.16 $&$ 1.03 $ &        $ 0.5 $&$ 1.08$&      $ 1.9 $&$ 0.806 $ &
$ 7. $&$ 0.11 $& $ 36 $&$ 1.52\, \times \, {10}^{-3} $&     $ 192
$&$ 1.13\, \times \, {10}^{-5} $ \cr $ 0.17 $&$ 1.04 $ &        $
0.55 $&$ 1.08 $ & $ 2. $&$ 0.776 $ &    $ 7.5 $&$ 9.45\, \times \,
{10}^{-2} $ & $40$&$ 1.12\, \times \, {10}^{-3} $ &    $208$&$
8.91\, \times \, {10}^{-6} $ \cr $ 0.18 $&$1.04$ &          $ 0.6
$&$ 1.08 $ & $2.1$&$ 0.746 $ &      $ 8. $&$8.15\,\times \,
{10}^{-2} $ &     $ 44 $&$ 8.55\, \times\,{10}^{-4} $ &      $ 224
$&$ 7.16\, \times\,{10}^{-6} $ \cr $ 0.19 $&$ 1.04 $ &        $ 0.65
$&$ 1.08$&     $ 2.2 $&$ 0.716 $ &    $ 9 $&$ 6.18\, \times
\,{10}^{-2}$ &      $ 48 $&$ 6.66\, \times \, {10}^{-4} $ &    $ 240
$&$ 5.82\, \times \, {10}^{-6} $ \cr $ 0.2 $&$ 1.04 $ & $ 0.7 $&$
1.08 $ &    $ 2.3 $&$ 0.687 $ &    $ 10 $&$ 4.79\, \times \,
{10}^{-2} $ &   $52$&$5.28\, \times \, {10}^{-4} $ & $256 $&$ 4.8\,
\times \, {10}^{-6} $ \cr $ 0.22$&$1.04 $ & $ 0.75 $&$ 1.08 $ & $2.4
$&$ 0.659 $ &    $ 11 $&$ 3.78\, \times \, {10}^{-2} $ &   $ 56 $&$
4.26\, \times\,{10}^{-4} $ & $ 288 $&$ 3.38\, \times\,{10}^{-6} $
\cr $ 0.24 $&$ 1.05 $ & $ 0.8 $&$ 1.07$& $ 2.5 $&$ 0.631 $ &    $ 12
$&$ 3.04\, \times\,{10}^{-2}$ &      $ 60 $&$ 3.49\, \times \,
{10}^{-4} $ & $ 320 $&$ 2.47\, \times \, {10}^{-6} $ \cr $ 0.26 $&$
1.05 $ & $ 0.85 $&$ 1.07 $ &   $ 2.6 $&$ 0.605 $&     $ 13 $&$
2.47\, \times \, {10}^{-2} $ &   $64$&$ 2.89\, \times \, {10}^{-4} $
& $352 $&$ 1.86\, \times \, {10}^{-6} $ \cr $0.28$&$ 1.05 $ & $ 0.9
$&$ 1.06 $ &   $ 2.7 $&$ 0.579 $ & $14$&$ 2.04\, \times \, {10}^{-2}
$ &     $ 72 $&$ 2.05\, \times\,{10}^{-4} $ &      $ 384
$&$1.43\,\times\, {10}^{-6} $ \cr
  $ 0.3 $&$ 1.06 $ &       $ 0.95 $&$ 1.05$&     $ 2.8 $&$ 0.554 $ &    $ 15 $&$ 1.7\, \times\,{10}^{-2} $ &      $ 80 $&$ 1.5\, \times \, {10}^{-4} $ &     $ 416 $&$ 1.13\, \times\,{10}^{-6}$ \cr
  $ 0.32 $&$ 1.06 $ &      $ 1. $&$ 1.05 $ &     $ 2.9 $&$ 0.531$&      $ 16 $&$ 1.43\, \times \, {10}^{-2} $ &   $ 88$&$1.14\, \times \, {10}^{-4} $ &      $ 448 $&$ 9.04\, \times \, {10}^{-7} $ \cr
 $0.34 $&$ 1.06 $ &        $ 1.1 $&$ 1.03 $ &    $ 3. $&$ 0.508 $ &     $18$&$1.05\, \times \, {10}^{-2} $ &      $ 96 $&$ 8.8\, \times\,{10}^{-5} $ &       $480$&$ 7.36\, \times \, {10}^{-7} $ \cr
  $ 0.36 $&$ 1.06 $ &      $ 1.2 $&$ 1.01$&      $ 3.5 $&$ 0.409 $ &    $ 20 $&$ 7.86\, \times\,{10}^{-3} $ &     $ 104 $&$ 6.95\, \times \, {10}^{-5} $ &   $ 512 $&$ 6.07\,\times\, {10}^{-7} $ \cr
  $ 0.38 $&$ 1.07 $ &     $ 1.3 $&$ 0.981 $ &    $ 4. $&$ 0.331$&       $ 22 $&$ 6.05\, \times \, {10}^{-3} $ &  $112$&$ 5.58\, \times \, {10}^{-5} $ & $1000$&  $8.18\,\times\, {10}^{-8} $\cr
 $0.4 $&$ 1.07 $ &         $ 1.4 $&$ 0.955 $ &   $ 4.5 $&$ 0.27 $ &     $24$&$ 4.76\, \times \, {10}^{-3} $ &     $ 120 $&$ 4.55\, \times\, {10}^{-5} $ & $2000$&$1.03\,\times\, {10}^{-8}$\cr
 \hline
 \end{tabular}
\caption{Function $F(x)/F(0)$ versus $x=q/m$; $F(0)=7/1152$.} \label{tableF}
\end{table}
\begin{table}[h]
\begin{tabular}{|c|c|c|c|c|c|c|}
\hline  & \multicolumn{4}{|c|}{$1S_{1/2}$} & $2S_{1/2}$ &
$2P_{1/2}$ \cr\hline
$Z$& $\Delta g_0+\Delta g_1$ & $\Delta g_{nr}$ & $\Delta g$ &
$\Delta g (\mbox{Ref. \cite{Beier2000}})$ & $8\Delta g$ & $8\Delta
g$ \cr\hline&&&&&&\cr
$ 1 $ & $ 4.935\times{10}^{-9} $ & $ 4.934\times{10}^{-9} $ & $
4.934\times{10}^{-9} $ &       & $ 4.936\times{10}^{-9} $ & $
1.638\times{10}^{-9} $ \cr
 $ 2 $ & $ 1.58\times{10}^{-7} $ & $ 1.58\times{10}^{-7} $ & $ 1.58\times{10}^{-7} $ &       & $ 1.58\times{10}^{-7} $ & $ 5.26\times{10}^{-8} $ \cr
 $ 3 $ & $ 1.2\times{10}^{-6} $ & $ 1.2\times{10}^{-6} $ & $ 1.2\times{10}^{-6} $ &       & $ 1.2\times{10}^{-6} $ & $ 4.01\times{10}^{-7} $ \cr
 $ 4 $ & $ 5.04\times{10}^{-6} $ & $ 5.04\times{10}^{-6} $ & $ 5.04\times{10}^{-6} $ &       & $ 5.05\times{10}^{-6} $ & $ 1.69\times{10}^{-6} $ \cr
 $ 5 $ & $ 1.53\times{10}^{-5} $ & $ 1.53\times{10}^{-5} $ & $ 1.54\times{10}^{-5} $ &       & $ 1.54\times{10}^{-5} $ & $ 5.18\times{10}^{-6} $ \cr
 $ 6 $ & $ 3.79\times{10}^{-5} $ & $ 3.8\times{10}^{-5} $ & $ 3.81\times{10}^{-5} $ &       & $ 3.82\times{10}^{-5} $ & $ 1.29\times{10}^{-5} $ \cr
 $ 7 $ & $ 8.16\times{10}^{-5} $ & $ 8.17\times{10}^{-5} $ & $ 8.2\times{10}^{-5} $ &       & $ 8.23\times{10}^{-5} $ & $ 2.81\times{10}^{-5} $ \cr
 $ 8 $ & $ 1.58\times{10}^{-4} $ & $ 1.58\times{10}^{-4} $ & $ 1.59\times{10}^{-4} $ &       & $ 1.6\times{10}^{-4} $ & $ 5.49\times{10}^{-5} $ \cr
 $ 9 $ & $ 2.83\times{10}^{-4} $ & $ 2.84\times{10}^{-4} $ & $ 2.86\times{10}^{-4} $ &       & $ 2.87\times{10}^{-4} $ & $ 9.94\times{10}^{-5} $ \cr
 $ 10 $ & $ 4.76\times{10}^{-4} $ & $ 4.78\times{10}^{-4} $ & $ 4.82\times{10}^{-4} $ &       & $ 4.84\times{10}^{-4} $ & $ 1.69\times{10}^{-4} $ \cr
 $ 11 $ & $ 7.61\times{10}^{-4} $ & $ 7.66\times{10}^{-4} $ & $ 7.72\times{10}^{-4} $ & $ 3(3)\times{10}^{-4} $ & $ 7.76\times{10}^{-4} $ & $ 2.73\times{10}^{-4} $ \cr
 $ 12 $ & $ 1.17\times{10}^{-3} $ & $ 1.18\times{10}^{-3} $ & $ 1.19\times{10}^{-3} $ & $ 4(5)\times{10}^{-4} $ & $ 1.19\times{10}^{-3} $ & $ 4.24\times{10}^{-4} $ \cr
 $ 13 $ & $ 1.72\times{10}^{-3} $ & $ 1.74\times{10}^{-3} $ & $ 1.76\times{10}^{-3} $ & $ 8(5)\times{10}^{-4} $ & $ 1.77\times{10}^{-3} $ & $ 6.35\times{10}^{-4} $ \cr
 $ 14 $ & $ 2.48\times{10}^{-3} $ & $ 2.51\times{10}^{-3} $ & $ 2.54\times{10}^{-3} $ & $ 1.4(1.0)\times{10}^{-3} $ & $ 2.56\times{10}^{-3} $ & $ 9.25\times{10}^{-4} $ \cr
 $ 15 $ & $ 3.46\times{10}^{-3} $ & $ 3.52\times{10}^{-3} $ & $ 3.57\times{10}^{-3} $ & $ 2(1)\times{10}^{-3} $ & $ 3.6\times{10}^{-3} $ & $ 1.31\times{10}^{-3} $ \cr
 $ 16 $ & $ 4.74\times{10}^{-3} $ & $ 4.82\times{10}^{-3} $ & $ 4.9\times{10}^{-3} $ & $ 3(1)\times{10}^{-3} $ & $ 4.95\times{10}^{-3} $ & $ 1.82\times{10}^{-3} $ \cr
 $ 17 $ & $ 6.35\times{10}^{-3} $ & $ 6.48\times{10}^{-3} $ & $ 6.6\times{10}^{-3} $ & $ 5(2)\times{10}^{-3} $ & $ 6.67\times{10}^{-3} $ & $ 2.48\times{10}^{-3} $ \cr
 $ 18 $ & $ 8.36\times{10}^{-3} $ & $ 8.56\times{10}^{-3} $ & $ 8.73\times{10}^{-3} $ & $ 6(2)\times{10}^{-3} $ & $ 8.83\times{10}^{-3} $ & $ 3.31\times{10}^{-3} $ \cr
 $ 20 $ & $ 1.39\times{10}^{-2} $ & $ 1.43\times{10}^{-2} $ & $ 1.46\times{10}^{-2} $ & $ 1.0(3)\times{10}^{-2} $ & $ 1.48\times{10}^{-2} $ & $ 5.66\times{10}^{-3} $ \cr
 $ 24 $ & $ 3.28\times{10}^{-2} $ & $ 3.45\times{10}^{-2} $ & $ 3.56\times{10}^{-2} $ & $ 3.3(3)\times{10}^{-2} $ & $ 3.63\times{10}^{-2} $ & $ 1.44\times{10}^{-2} $ \cr
 $ 28 $ & $ 6.72\times{10}^{-2} $ & $ 7.22\times{10}^{-2} $ & $ 7.53\times{10}^{-2} $ & $ 6.9(3)\times{10}^{-2} $ & $ 7.7\times{10}^{-2} $ & $ 3.19\times{10}^{-2} $ \cr
 $ 32 $ & $ 0.123 $ & $ 0.136 $ & $ 0.144 $ & $ 0.138 $ & $ 0.148 $ & $ 6.37\times{10}^{-2} $ \cr
 $ 36 $ & $ 0.207 $ & $ 0.238 $ & $ 0.254 $ & $ 0.249 $ & $ 0.262 $ & $ 0.118 $ \cr
 $ 40 $ & $ 0.325 $ & $ 0.389 $ & $ 0.421 $ & $ 0.410 $ & $ 0.437 $ & $ 0.206 $ \cr
 $ 44 $ & $ 0.481 $ & $ 0.607 $ & $ 0.665 $ & $ 0.658 $ & $ 0.695 $ & $ 0.341 $ \cr
 $ 48 $ & $ 0.676 $ & $ 0.907 $ & $ 1.01 $ & $ 1.01 $ & $ 1.06 $ & $ 0.545 $ \cr
 $ 52 $ & $ 0.904 $ & $ 1.31 $ & $ 1.48 $ & $ 1.48 $ & $ 1.56 $ & $ 0.841 $ \cr
 $ 56 $ & $ 1.15 $ & $ 1.84 $ & $ 2.1 $ & $ 2.12 $ & $ 2.24 $ & $ 1.26 $ \cr
 $ 60 $ & $ 1.41 $ & $ 2.51 $ & $ 2.92 $ & $ 2.95 $ & $ 3.13 $ & $ 1.85 $ \cr
 $ 64 $ & $ 1.63 $ & $ 3.35 $ & $ 3.97 $ & $ 4.03 $ & $ 4.29 $ & $ 2.66 $ \cr
 $ 68 $ & $ 1.77 $ & $ 4.4 $ & $ 5.3 $ & $ 5.39 $ & $ 5.77 $ & $ 3.76 $ \cr
 $ 72 $ & $ 1.78 $ & $ 5.67 $ & $ 6.96 $ & $ 7.11 $ & $ 7.62 $ & $ 5.23 $ \cr
 $ 76 $ & $ 1.55 $ & $ 7.2 $ & $ 9. $ & $ 9.24 $ & $ 9.93 $ & $ 7.18 $ \cr
 $ 80 $ & $ 0.983 $ & $ 9.02 $ & $ 11.5 $ & $ 11.9 $ & $ 12.8 $ & $ 9.75 $ \cr
 $ 83 $ & $ 0.252 $ & $ 10.6 $ & $ 13.7 $ & $ 14.2 $ & $ 15.3 $ & $ 12.2 $ \cr
 $ 88 $ & $ -1.77 $ & $ 13.7 $ & $ 18.1 $ & $ 18.9 $ & $ 20.5 $ & $ 17.5 $ \cr
 $ 92 $ & $ -4.34 $ & $ 16.5 $ & $ 22.5 $ & $ 23.5 $ & $ 25.5 $ & $ 23.1 $ \cr
 \hline
\end{tabular}
 \caption{The quantity $n^3\Delta g$ in units
$10^{-6}$, calculated in various approximations for $1S_{1/2}$,
$2S_{1/2}$, and $2P_{1/2}$ states. Our results are obtained with the
account for the magnetic loop in the leading approximation
(contribution of light-by-light scattering). The quantity $\Delta
g_{nr}$ denotes the correction obtained with the use of Eq.
(\ref{eq:TM1f}) with the functions $f_1(r)$ and $f_2(r)$ taken in
the nonrelativistic approximation, see Eq. (\ref{eq:fg_nr}). }
\label{table1}
\end{table}
\begin{table}[h]
\begin{tabular}{|c|c|c|c|c|c|}
\hline%
$ Z  $   &    $\Delta g              $   &   $\Delta g_{as}$&$ Z  $   &   $\Delta g              $   &   $\Delta g_{as}$ \cr
\hline&&&&&\cr
$ 1  $   &    $ 1.043\times{10}^{-2} $   &   $-0.2701$   &   $ 24 $   &   $ 159.5                $   &   $ 145.9 $   \cr
$ 2  $   &    $ 0.1274               $   &   $-0.6226$   &   $ 28 $   &   $ 233.7                $   &   $ 217.9 $   \cr
$ 3  $   &    $ 0.484                $   &   $-0.7983$   &   $ 32 $   &   $ 324.                 $   &   $ 306.1 $   \cr
$ 4  $   &    $ 1.186                $   &   $-0.6592$   &   $ 36 $   &   $ 430.8                $   &   $ 411.  $   \cr
$ 5  $   &    $ 2.317                $   &   $-0.109 $   &   $ 40 $   &   $ 554.4                $   &   $ 532.8 $   \cr
$ 6  $   &    $ 3.944                $   &   $ 0.9268$   &   $ 44 $   &   $ 694.9                $   &   $ 671.5 $   \cr
$ 7  $   &    $ 6.124                $   &   $ 2.508 $   &   $ 48 $   &   $ 852.3                $   &   $ 827.4 $   \cr
$ 8  $   &    $ 8.904                $   &   $ 4.687 $   &   $ 52 $   &   $ 1026.                $   &   $ 1000. $   \cr
$ 9  $   &    $ 12.33                $   &   $ 7.506 $   &   $ 56 $   &   $ 1217.                $   &   $ 1189. $   \cr
$ 10 $   &    $ 16.43                $   &   $ 11.   $   &   $ 60 $   &   $ 1424.                $   &   $ 1395. $   \cr
$ 11 $   &    $ 21.24                $   &   $ 15.22 $   &   $ 64 $   &   $ 1646.                $   &   $ 1616. $   \cr
$ 12 $   &    $ 26.8                 $   &   $ 20.17 $   &   $ 68 $   &   $ 1883.                $   &   $ 1852. $   \cr
$ 13 $   &    $ 33.13                $   &   $ 25.9  $   &   $ 72 $   &   $ 2134.                $   &   $ 2103. $   \cr
$ 14 $   &    $ 40.26                $   &   $ 32.42 $   &   $ 76 $   &   $ 2398.                $   &   $ 2366. $   \cr
$ 15 $   &    $ 48.2                 $   &   $ 39.77 $   &   $ 80 $   &   $ 2673.                $   &   $ 2641. $   \cr
$ 16 $   &    $  56.98               $   &   $ 47.96 $   &   $ 83 $   &   $ 2886.                $   &   $ 2854. $   \cr
$ 17 $   &    $ 66.62                $   &   $ 57.01 $   &   $ 88 $   &   $ 3251.                $   &   $ 3219. $   \cr
$ 18 $   &    $ 77.14                $   &   $ 66.94 $   &   $ 90 $   &   $ 3400.                $   &   $ 3368. $   \cr
$ 20 $   &    $ 100.9                $   &   $ 89.52 $   &   $ 92 $   &   $ 3550.                $   &   $ 3519. $   \cr
\hline
\end{tabular}
 \caption{$\Delta g$ in units $10^{-6}$ for $1S_{1/2}$ state
of muonic atom. $\Delta g_{as}$ is the asymptotics
(\ref{eq:mu_as}).}
 \label{tablemu}
\end{table}
\end{document}